\documentclass[12pt]{article}
\usepackage{amsmath,amssymb,amsthm,amsxtra,bbm,overpic,bm,epsfig,subfigure}
\usepackage[colorlinks]{hyperref}
\usepackage{mathrsfs}
\usepackage{enumitem}
\usepackage{graphicx}
\usepackage{color}
\usepackage{comment}
\usepackage{epstopdf}
\usepackage{float}
\usepackage{cite}
\usepackage{upgreek,overpic,setspace,multirow,textcomp,orcidlink}
\numberwithin{equation}{section}
\allowdisplaybreaks[2]

\usepackage{slashed,stmaryrd}
\usepackage{subcaption}

\def\thefootnote{\fnsymbol{footnote}}

\usepackage{lscape}%
\usepackage{array}
\usepackage{booktabs}%

\begin{document}
	
	\vspace{0.2cm}
	
	\begin{center}
        {\Large\bf One-Loop Effects in the Neutrino Matter Potential\\ and Implications for Non-Standard Interactions}
    \end{center}
	
	\vspace{0.2cm}
	
	\begin{center}
		{\bf Jihong Huang}{\orcidlink{0000-0002-5092-7002}},$^{1,2}$\footnote{E-mail: huangjh@ihep.ac.cn}
		\ 
		{\bf Tommy Ohlsson}{\orcidlink{0000-0002-3525-8349}},$^{3,4}$\footnote{E-mail: tohlsson@kth.se}
		\ 
		{\bf Sampsa Vihonen}{\orcidlink{0000-0001-7761-2847}},$^{3,4}$\footnote{E-mail: vihonen@kth.se}
		\ 
		{\bf Shun Zhou}{\orcidlink{0000-0003-4572-9666}},$^{1,2}$\footnote{E-mail: zhoush@ihep.ac.cn}
		\\
		\vspace{0.2cm}
		{\small
			$^{1}$Institute of High Energy Physics, Chinese Academy of Sciences, Beijing 100049, China\\
			$^{2}$School of Physical Sciences, University of Chinese Academy of Sciences, Beijing 100049, China\\
			$^{3}$Department of Physics, School of Engineering Sciences, KTH Royal Institute of Technology,\\
			AlbaNova University Center, Roslagstullsbacken 21, SE–106 91 Stockholm, Sweden\\
			$^{4}$The Oskar Klein Centre, AlbaNova University Center, Roslagstullsbacken 21,\\
			SE–106 91 Stockholm, Sweden}
	\end{center}

	\vspace{0.5cm}
	
	\begin{abstract}
		In this work, we emphasize that it is necessary to take into account one-loop corrections of $2.0\%$ to the neutrino matter potential in the precision measurements of neutrino oscillation parameters and in the experimental searches for new physics beyond the Standard Model. With the numerical simulation of the DUNE experiment, we study how radiative corrections to the matter potential affect neutrino oscillation probabilities, and thus, the event rates in the presence of neutrino non-standard interactions (NSIs). We find that neglecting one-loop corrections may lead to wrong conclusions for the discovery of NSIs. The implications for the determination of neutrino mass ordering and constraints on the NSI parameters in future long-baseline accelerator neutrino experiments are explored in a quantitative way. 
		
	\end{abstract}
	
	\def\thefootnote{\arabic{footnote}}
	\setcounter{footnote}{0}
	
	\newpage
	
		\section{Introduction}
	
	\label{sec:introduction}
	
	Neutrino oscillation experiments over the past few decades have provided us with robust evidence that neutrinos are massive and leptonic flavor mixing is significant~\cite{ParticleDataGroup:2024cfk,Xing:2020ijf}. Various extensions of the Standard Model (SM), which have been proposed to understand the origin of neutrino masses and leptonic flavor mixing, may also induce beyond-the-SM (BSM) effects in neutrino oscillation phenomenology. One appealing case is to consider neutrino non-standard interactions (NSIs) with ordinary matter described by the following effective Lagrangian
	\begin{eqnarray}
		\label{eq:LNSI}
		{\cal L}_{\rm NSI}^{m} = -2\sqrt{2} G_\mu^{} \epsilon_{\alpha \beta}^{f {\rm C}} \left( \overline{\nu_\alpha^{}} \gamma^\mu P_{\rm L}^{} \nu_\beta^{}\right) \left( \overline{f} \gamma_\mu P_{\rm C}^{} f\right) \;, 
	\end{eqnarray}
	where $\alpha,\beta = e,\mu,\tau$ refer to three lepton flavors and $f=e,u,d$ denotes the first-generation charged fermions in the SM. In addition, $G_\mu^{} \approx 1.166 \times 10^{-5}~{\rm GeV}^{-2}$ stands for the Fermi constant, $P_{\rm C}^{}$ for ${\rm C} = {\rm L}$ (or ${\rm C} = {\rm R}$) is the left-handed (or right-handed) chiral projection operator with $P^{}_{\rm L, R} \equiv \left(1\mp \gamma^5 \right)/2$, and $\epsilon_{\alpha \beta}^{f {\rm C}}$ (for ${\rm C} = {\rm L, R}$) are the NSI parameters. Such NSIs can also affect neutrino flavor oscillations in ordinary matter, similar to the standard matter effects caused by coherent forward scattering between neutrinos and background particles~\cite{Wolfenstein:1977ue,Wolfenstein:1979ni,Mikheyev:1985zog,Mikheev:1986wj}. See, e.g., Refs.~\cite{Ohlsson:2012kf,Farzan:2017xzy}, for recent reviews on the NSIs and related phenomenological implications. Generally speaking, those four-fermion operators in Eq.~(\ref{eq:LNSI}) can be regarded as a subset of dimension-six operators in the framework of the SM effective field theory with the corresponding Wilson coefficients $C_{\alpha\beta f f {\rm C}}^{(6)}$~\cite{Brivio:2017vri}. In this case, the NSI parameters are related to the Wilson coefficients by $\epsilon_{\alpha \beta}^{f {\rm C}} \propto \left(m_W^2/\Lambda^2\right) C_{\alpha\beta f f {\rm C}}^{(6)}$, where $m_W^{}$ and $\Lambda$ are the $W$-boson mass and the cutoff scale, respectively.
	
	The primary goals for the next generation of neutrino oscillation experiments are to determine whether three neutrino masses $m_1^{},m_2^{}$ and $m_3^{}$ take on the normal ordering (NO, i.e., $m_1^{}<m_2^{}<m_3^{}$) or the inverted ordering (IO, i.e., $m_3^{}<m_1^{}<m_2^{}$), and to discover the leptonic CP violation. Meanwhile, neutrino oscillation parameters will be measured with sub-percent precisions~\cite{JUNO:2022mxj,Capozzi:2025wyn}. To match the future experimental precisions, we shall take into account electroweak one-loop effects in the SM, which are normally at the percent level. This is important both for extracting neutrino oscillation parameters and for constraining BSM physics effects. Radiative corrections to neutrino interactions in the SM have already been calculated in detail, and their experimental impacts have been extensively discussed in Refs.~\cite{Bahcall:1995mm,Kurylov:2001av,Kurylov:2002vj,Marciano:2003eq,Tomalak:2019ibg,Tomalak:2020zfh,Huang:2024rfb}. On the other hand, the matter effects on neutrino oscillations, described by the matter potential~\cite{Wolfenstein:1977ue}, play an important role in long-baseline accelerator neutrino experiments, such as DUNE~\cite{DUNE:2020ypp}. The one-loop corrections to the matter potential have been calculated in Refs.~\cite{Botella:1986wy,Mirizzi:2009td,Huang:2023nqf}, where a $2.0\%$ correction to the tree-level charged-current (CC) potential has been found, and its impact on future long-baseline experiments has been studied in Ref.~\cite{Huang:2025apv}.
	
	In this work, we continue to explore the impact of one-loop corrections to the matter potential on long-baseline accelerator neutrino experiments, which are expected to discover or constrain BSM effects with precision data. To be explicit, we take the BSM effects to be the NSIs given in Eq.~(\ref{eq:LNSI}). First, starting from the effective Hamiltonian for neutrino oscillations in matter, we derive the oscillation probabilities and highlight the deviations from the standard results caused by both radiative corrections and the NSIs. Then, the oscillation probabilities of the appearance channel $\nu_\mu^{} \to \nu_e^{}$ at DUNE are simulated for both neutrinos and antineutrinos. The differences from those in the standard case are illustrated and the mimicking effects of NSIs are also pointed out. Finally, the influence on determining the neutrino mass ordering is analyzed. By varying the off-diagonal NSI parameter $\epsilon_{e \mu}^{}$, we find that the sensitivity of DUNE to neutrino mass ordering decreases by about $4.2\sigma$ confidence level (CL), while radiative corrections to the matter potential will slightly increase it. 
	
	The remaining part of this work is organized as follows. In Sec.~\ref{sec:NSI}, we derive the oscillation probabilities and discuss the corrections from both one-loop effects and NSIs. With the simulation methods introduced in Sec.~\ref{sec:methods}, we perform the numerical analyses in Sec.~\ref{sec:results}, including the experimental constraints on the NSI parameters and the sensitivity to the determination of neutrino mass ordering. Finally, in Sec.~\ref{sec:sum}, our main results and conclusions are summarized.

    	\section{Neutrino Non-Standard Interactions}
	
	\label{sec:NSI}

	After adding the interaction Lagrangian in Eq.~(\ref{eq:LNSI}) into the SM, one can immediately recognize that the time evolution of neutrino flavor eigenstates in matter is described by the Schr\"{o}dinger-like equation with an effective Hamiltonian
	\begin{eqnarray}
		\label{eq:NSI_Heff}
		{\rm i} \frac{{\rm d}}{{\rm d} t} \begin{pmatrix} \nu_e^{} \\ \nu_\mu^{} \\ \nu_\tau^{} \end{pmatrix} = \frac{1}{2 E_{\nu}^{}} \left[ U
		\begin{pmatrix}
			0 & 0 & 0 \\
			0 & \Delta m_{21}^2 & 0 \\
			0 & 0 & \Delta m_{31}^2
		\end{pmatrix} U^{\dagger}
		+ A
		\begin{pmatrix}
			1 + \epsilon_{ee}^{} & \epsilon_{e\mu}^{} & \epsilon_{e\tau}^{} \\
			\epsilon_{e\mu}^* & \epsilon_{\mu\mu}^{} & \epsilon_{\mu\tau}^{} \\
			\epsilon_{e\tau}^* & \epsilon_{\mu\tau}^* & \epsilon_{\tau\tau}^{}
		\end{pmatrix} \right] 
		\begin{pmatrix} \nu_e^{} \\ \nu_\mu^{} \\ \nu_\tau^{} \end{pmatrix} \;,
	\end{eqnarray}
	where $E^{}_\nu$ is the neutrino energy. The first matrix in the square brackets on the right-hand side of Eq.~(\ref{eq:NSI_Heff}) is determined by two neutrino mass-squared differences $\Delta m_{21}^2 \equiv m_2^2 - m_1^2$ and $\Delta m_{31}^2 \equiv m_3^2 - m_1^2$ and the Pontecorvo-Maki-Nakagawa-Sakata matrix $U$~\cite{Pontecorvo:1957cp,Maki:1962mu,Pontecorvo:1967fh}, which contains three mixing angles $\{\theta_{12}^{},\theta_{13}^{},\theta_{23}^{}\}$ and one Dirac CP-violating phase $\delta_{\rm CP}^{}$ in the standard parametrization~\cite{ParticleDataGroup:2024cfk}. The matter effects induced by the standard CC interaction and NSIs are represented by the second matrix in the square brackets, where the parameter $A = 2 \sqrt{2} G_\mu^{} N_e^{} E_\nu^{}$ characterizes the tree-level matter effect arising from the CC interaction with electrons in the medium, for which the electron number density is $N_e^{}$. On the other hand, the {\it effective} matter NSI parameters $\epsilon^{}_{\alpha \beta}$ (for $\alpha, \beta = e, \mu, \tau$) are defined as
	\begin{eqnarray}
		\epsilon_{\alpha \beta}^{} \equiv \sum_{f,{\rm C}} \epsilon_{\alpha \beta}^{f {\rm C}} \frac{N_f^{}}{N_e^{}} \;,
	\end{eqnarray}
	with the number density $N_f^{}$ for $f=e,u,d$ in the medium. Notice that the identity $\epsilon_{\alpha \beta}^{} = \epsilon_{\beta \alpha}^*$ must hold due to the Hermiticity of the effective Hamiltonian. One may further rewrite the complex off-diagonal elements as $\epsilon_{\alpha \beta}^{} \equiv \left|\epsilon_{\alpha \beta}^{}\right| \exp ({\rm i} \phi_{\alpha\beta}^{})$ (for $\alpha \neq \beta$). Therefore, there are eight independent real parameters in the matter NSIs, i.e., three amplitudes and three phases of off-diagonal parameters, and two amplitudes of diagonal elements after subtracting an overall constant multiple of identity. A recent global-fit analysis of neutrino oscillation data places the following constraints on the individual NSI parameters: $|\epsilon_{e\mu}^{}| \in [-0.32, 0.40], |\epsilon_{e\tau}^{}| \in [-0.49, 0.45], |\epsilon_{\mu\tau}^{}| \in [-0.043, 0.039],$ and also $\epsilon_{ee}^{} - \epsilon_{\mu\mu}^{} \in [-4.8,-1.6] \oplus [-0.40, 2.6]$ and $\epsilon_{\tau\tau}^{} - \epsilon_{\mu\mu}^{} \in [-0.075, 0.080]$ ($99\%$~CL)~\cite{Coloma:2023ixt}.
    
    In long-baseline accelerator neutrino experiments, the appearance channel $\nu_{\mu}^{} \to \nu_{e}^{}$ with its oscillation probability $P_{\mu e} \equiv P\left(\nu_{\mu}^{} \to \nu_{e}^{}\right)$ and its CP-conjugate channel $\overline{\nu}_{\mu}^{} \to \overline{\nu}_{e}^{}$ with $\overline{P}_{\mu e}^{} \equiv P\left(\overline{\nu}_{\mu}^{} \to \overline{\nu}_{e}^{}\right)$ will be used to determine the neutrino mass ordering and measure the CP-violating phase $\delta_{\rm CP}^{}$. In the standard case, with series expansions in terms of small parameters $\Delta m^2_{21}/\Delta m^2_{31}$ and $\sin^2 \theta^{}_{13}$, the oscillation probability $P_{\mu e}^0$ can be approximately written as~\cite{Cervera:2000kp,Freund:2001pn,Akhmedov:2004ny,Nunokawa:2007qh}
	\begin{eqnarray}
        \label{eq:P0}
		P_{\mu e}^{0} &\simeq& \sin^2\theta_{23}^{}  \sin^2 2 \theta_{13}^{}  \frac{\sin^2\left(\Delta_{31}^{} - a L \right)}{\left(\Delta_{31}-a L\right)^2} \Delta_{31}^2  \nonumber \\
		&& + \sin2\theta_{23}^{} \sin2\theta_{13}^{} \sin2\theta_{12}^{} \frac{\sin \left(\Delta_{31}^{} - a L\right)}{\left(\Delta_{31}^{}-a L\right)}  \frac{\sin(a L)}{(a L)} \Delta_{31}^{} \Delta_{21}^{} \cos\left(\Delta_{31}^{}+\delta_{\rm CP}^{} \right)  \nonumber \\
		&& + \cos^2\theta_{23}^{} \sin^2 2\theta_{12}^{} \frac{\sin^2(a L)}{(a L)^2} \Delta_{21}^2 \;,
	\end{eqnarray}
	with the baseline length $L$, the oscillation phase $\Delta_{ij}^{} \equiv \Delta m_{ij}^2 L /(4E_\nu^{})$ and the tree-level matter parameter $a \equiv A/(4 E_\nu^{})$. 
    
    At one-loop level, the $2.0\%$ correction to the CC matter potential can be incorporated into the effective Hamiltonian by replacing the $(e,e)$-element ``$1+\epsilon^{}_{ee}$" of the second matrix to ``$1.020+\epsilon^{}_{ee}$", while the neutral-current (NC) potential is universal for all three types of neutrinos, and thus, it is irrelevant for subsequent discussions. Based on the oscillation probability in Eq.~(\ref{eq:P0}) and the one-loop matter parameter $\overline{a} \simeq a (1+\delta a)$ with $\delta a = 2.0\%$ being a constant~\cite{Huang:2023nqf}, extra contributions to the oscillation probability from one-loop corrections are given by 
	\begin{eqnarray}
    \label{eq:Ploop}
		\Delta P_{\mu e}^{\rm loop} &\simeq& 2 \delta a (a L) \sin^2\theta_{23}^{} \sin^2 2\theta_{13}^{}  \left[\frac{\sin^2\left(\Delta_{31}^{}-a L\right)}{\left(\Delta_{31}^{}-a L\right)^3}-\frac{\sin2\left(\Delta_{31}^{}-a L\right)}{2\left(\Delta_{31}^{}-a L\right)^2}\right] \Delta_{31}^2  \nonumber \\
		&& + \delta a \sin2\theta_{23}^{} \sin2\theta_{13}^{} \sin2\theta_{12}^{}  \left[\frac{\sin\left(\Delta_{31}^{} - a L\right)}{\left(\Delta_{31}^{}-a L\right)} \left(\cos(a L) - \frac{\sin(a L)}{(aL)} \right) \right. \nonumber \\
		&& \left. + \frac{\sin\left(\Delta_{31}^{} - a L\right)}{\left(\Delta_{31}^{}-a L\right)^2} \sin(a L)  - \frac{\cos \left(\Delta_{31}^{} - a L\right)}{\left(\Delta_{31}^{}-a L\right)} \sin (a L) \right] \Delta_{31}^{} \Delta_{21}^{} \cos\left(\Delta_{31}^{} + \delta_{\rm CP}^{} \right)  \nonumber \\
		&& + \delta a \cos^2\theta_{23}^{} \sin^2 2\theta_{12}^{}  \left[\frac{\sin (2a L) }{(a L)}-\frac{2 \sin^2(a L)}{(a L)^2}\right] \Delta_{21}^2 \;.
	\end{eqnarray}
	Meanwhile, given the effective Hamiltonian in Eq.~(\ref{eq:NSI_Heff}), one can first calculate the oscillation probabilities by using the effective neutrino masses $\widetilde{m}_i^{}$ and mixing matrix elements $\widetilde{U}_{\alpha i}^{}$ in matter with NSIs and then mapping effective parameters into those vacuum $\{m_i^{},U_{\alpha i}^{},A,\epsilon_{\alpha \beta}\}$~\cite{Kopp:2007ne,Meloni:2009ia}. In this way, one can find out the modifications to the oscillation probability
	\begin{eqnarray}
    \label{eq:PNSI}
		\Delta P_{\mu e}^{\rm NSI} &\simeq& 4 \left|\epsilon_{e\mu}^{}\right| \cos^2\theta_{23}^{} \sin2\theta_{13}^{} \sin\theta_{23}^{} \sin (a L) \frac{\sin \left(\Delta_{31}^{} - a L\right) }{\left(\Delta_{31}^{} - a L\right)} \Delta_{31}^{} \cos\left(\Delta_{31}^{}+\delta_{\rm CP}^{}+\phi_{e\mu}^{}\right) \nonumber\\
		&& + 4 \left|\epsilon_{e\mu}^{}\right| a L  \sin2\theta_{13}^{} \sin^3\theta_{23}^{}     \frac{\sin^2\left(\Delta_{31}^{} - a L\right)}{\left(\Delta_{31}^{}-a L\right)^2} \Delta_{31}^{} \cos \left(\delta_{\rm CP}^{}+\phi_{e\mu}^{}\right) \nonumber \\
		&& + 4 \left|\epsilon_{e\mu}^{}\right|  \cos^3\theta_{23}^{}  \sin 2\theta_{12}^{}  \frac{\sin^2(a L) }{a L} \Delta_{21} \cos\phi_{e\mu}^{} \nonumber \\
		&& + 4 \left|\epsilon_{e\mu}^{}\right| \cos\theta_{23}^{} \sin 2\theta_{12}^{} \sin^2\theta_{23}^{} \sin (a L) \frac{\sin \left(\Delta_{31}^{} - a L\right)}{\Delta_{31}^{}-a L} \Delta_{21}^{} \cos\left(\Delta_{31}^{}-\phi_{e\mu}^{}\right) \;,
	\end{eqnarray}
    where we have retained only the NSI parameter $\epsilon_{e\mu}^{}$ for illustration. Hence, the total oscillation probability can be rewritten as $P_{\mu e}^{} = P_{\mu e}^0 + \Delta P_{\mu e}^{\rm loop} + \Delta P_{\mu e}^{\rm NSI}$, in which the effects from both the one-loop corrections and NSIs can be easily identified. In the remainder of this work, we define $P_{\mu e}^{\rm NSI}$ and $P_{\mu e}^{\rm SI}$ as the probabilities that are computed with and without NSI effects, respectively. Moreover, one needs to make the replacements $\delta_{\rm CP}^{} \to - \delta_{\rm CP}^{}$, $\phi_{e\mu}^{} \to - \phi_{e\mu}^{} $ and $a \to -a$ in $P_{\mu e}^{}$ to obtain the oscillation probability $\overline{P}_{\mu e}^{}$ for antineutrinos.\footnote{The sign of $\delta a$ does not need to be changed since the relative corrections for neutrinos and antineutrinos are the same.} Some helpful comments are in order:
	\begin{itemize}
		\item The three terms in $P_{\mu e}^0$ and $\Delta P_{\mu e}^{\rm loop}$ are proportional to $\Delta_{31}^2$, $\Delta_{31}^{}\Delta_{21}^{}$ and $\Delta_{21}^{2}$, respectively, whereas two of four terms in $\Delta P_{\mu e}^{\rm NSI}$ are proportional to $\Delta_{21}^{}$ and the other two to $\Delta_{31}^{}$ in the first-order approximation. Furthermore, the new phase $\phi_{e\mu}^{}$ could modify the oscillation probabilities together with $\delta_{\rm CP}^{}$ and $\Delta_{31}^{}$. Therefore, it is actually very difficult to analytically examine the main features of the total corrections to the oscillation probability based on these two expressions.
		
		\begin{figure}[!t]
			\centering
            \includegraphics[width=0.7\linewidth]{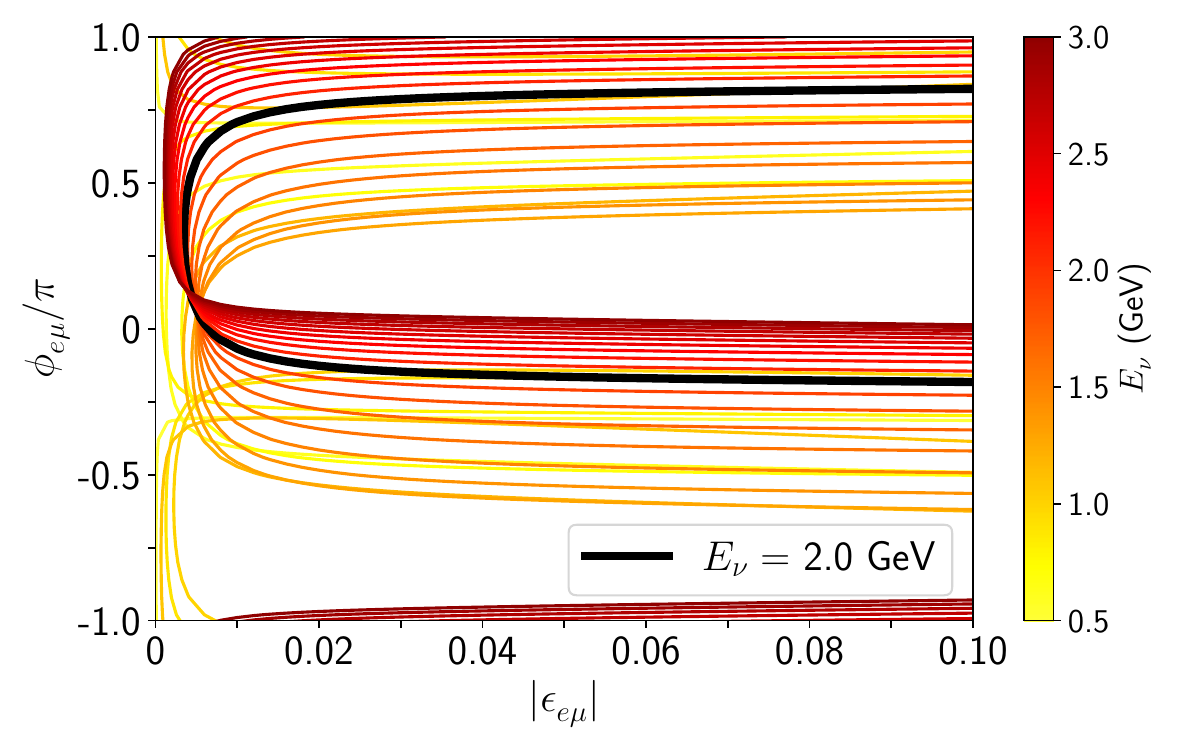}
			\caption{The values of $|\epsilon_{e\mu}^{}|$ and $\phi_{e\mu}^{}$ that allow the NSI corrections $\Delta P^{\rm NSI}_{\mu e}$ to reproduce $\Delta P^{\rm loop}_{\mu e}$ which arises from the one-loop corrections. When $\Delta P^{\rm NSI}_{\mu e} = \Delta P^{\rm loop}_{\mu e}$ holds, the NSI effects can successfully mimic the effect of one-loop corrections in neutrino matter potential. The probabilities are obtained numerically for the DUNE setup, where the baseline length is $L=1284.9~{\rm km}$, and a range of neutrino energies $E_\nu^{} \in [0.5, 3.0]$~GeV. The values of neutrino oscillation parameter are taken from Ref.~\cite{Esteban:2024eli} for normal ordering.}
			\label{fig:prob2}
		\end{figure}
		
		\item Using the exact oscillation probabilities in both cases, we numerically plot contours in the plane of the magnitude of the NSI parameter $\left|\epsilon_{e\mu}^{}
		\right|$ and its complex phase $\phi_{e\mu}^{}$ in Figure~\ref{fig:prob2}, in which NSIs can successfully {\it mimic} the matter effect at one-loop level, i.e., satisfying the relation $\Delta P_{\mu e}^{\rm loop} = \Delta P_{\mu e}^{\rm NSI}$ for each pair of $\{\left|\epsilon_{e\mu}^{}
		\right|,\phi_{e\mu}^{}\}$ along the curves. For illustration, we adopt the DUNE setup~\cite{DUNE:2021cuw} with the baseline length $L=1284.9~{\rm km}$ and the neutrino energy ranging from $0.5~{\rm GeV}$ (yellow curves) to $3.0~{\rm GeV}$ (red curves). Other oscillation parameters are fixed at their best-fit values from the latest global analysis of neutrino oscillation data in Ref.~\cite{Esteban:2024eli} in the NO case, and the one-loop correction to the matter potential is fixed as $\delta a = 2.0\%$. We notice the periodicity of $\phi_{e\mu}^{}$ from the plot, which comes from the fact that the NSI parameter $\epsilon_{e\mu}^{}$ and the oscillation probability in Eq.~(\ref{eq:PNSI}) keep unchanged for $\phi_{e\mu}^{} \to \phi_{e\mu}^{} + 2\pi$. In addition, for given values of the neutrino energy $E^{}_\nu$ and $\left|\epsilon_{e\mu}^{}
		\right|$, there are two values of $\phi_{e\mu}^{}$ solving the equation $\Delta P_{\mu e}^{\rm loop} = \Delta P_{\mu e}^{\rm NSI}$ in most cases. 
	\end{itemize}
	
	From the foregoing discussions on the one-loop effects and NSIs, we expect that there could be an interesting interplay between them. In the following, such an interplay will be quantitatively examined by simulating neutrino and antineutrino events in the DUNE experiment.

    \section{Simulation Methods}
    
    \label{sec:methods}

    For the simulation of the DUNE setup, we adopt the configuration that describes the experimental setup which was detailed in the DUNE Technical Design Report (TDR)~\cite{DUNE:2020ypp}. The simulation files are provided for \texttt{GLoBES} by the DUNE collaboration in Ref.~\cite{DUNE:2021cuw}. In this section, we provide a brief summary of the simulation methods that are used in this work.
    
    DUNE is a proposed next-generation long-baseline accelerator neutrino experiment currently under construction. The experimental setup comprises of the NuMI beamline in the Fermi National Accelerator Laboratory, which is currently undergoing an upgrade to a neutrino beam facility of 2.4~MW beam power. The far detector of DUNE is planned to consist of four modules of Liquid Argon Time Projection Chamber (LArTPC) design, which would have a combined fiducial mass of 40~kt. The far detector would be placed inside a mine at the Sanford Underground Research Facility, which is located about 1300~km from the planned neutrino source. DUNE is set to study neutrino oscillations by sending high-power beams of $\nu_\mu^{}$ and $\overline{\nu}_\mu^{}$ and observing oscillations in the $\nu_\mu^{} \rightarrow \nu_e^{}, \nu_\mu^{} \rightarrow \nu_\mu^{}, \overline{\nu}_\mu^{} \rightarrow \overline{\nu}_e^{}$ and $\overline{\nu}_\mu^{} \rightarrow \overline{\nu}_\mu^{}$ channels. According to DUNE TDR, the super-beam program is expected to run 6.5~years in $\nu_\mu^{}$ mode and 6.5~years in $\overline{\nu}_\mu^{}$ mode.

    In this work, the simulation of DUNE is done with the \texttt{GLoBES}~\cite{Huber:2004ka,Huber:2007ji} framework. \texttt{GLoBES} is a versatile simulation software that is used to study long-baseline neutrino oscillation experiments such as DUNE. \texttt{GLoBES} computes the expected number of neutrino events in each analysis bin and takes into account neutrino oscillations and provides a realistic approximation for the detector response. For the calculation of neutrino oscillation probabilities with the NSI effects, \texttt{GLoBES} is extended with the \texttt{snu} add-on. The matter density profile for DUNE is approximated with the well-known Shen-Ritzwoller profile~\cite{Roe:2017zdw}. The one-loop effects in the matter potential are included by adopting the methods that were introduced in Ref.~\cite{Huang:2025apv}.
    
    The statistical analysis of the simulated neutrino and antineutrino events is carried out with the chi-squared function,
    \begin{eqnarray}
        \label{eq:chi2}
            \chi_{}^2 = 2 \sum_{i=1}^{n} \left[ N_i^{\rm test} - N_i^{\rm true} + N_i^{\rm true} \ln \left(\frac{N_i^{\rm true}}{N_i^{\rm test}}\right) \right] + \sum_{k} \frac{\zeta_{k}^2}{\sigma_{k}^2} \;,
    \end{eqnarray}
    where the index $i$ runs through the neutrino energy bins $i = 1, 2, \ldots n$, where $n$ is the number of neutrino energy bins in the given channel. $N_{i}^{\rm true}$ and $N_{i}^{\rm test}$ are the simulated neutrino events that are calculated by \texttt{GLoBES} for the test values and the true values for the neutrino oscillation parameters. The evaluation of the systematic uncertainties is done with the well-known pull method. In Eq.~(\ref{eq:chi2}), the nuisance parameters $\zeta_{k}^{}$ parameterize the normalization errors assumed for signal and background events, whereas $\sigma_{k}$ are the corresponding uncertainties that are given at $1\sigma$~CL. The total number of events in each analysis bin is then computed as the sum of the corresponding signal and background events.

    The systematic uncertainties in DUNE are approximated with nine types of normalization errors~\cite{DUNE:2021cuw}. For the $\nu_\mu^{} \rightarrow \nu_e^{}$ channel, the signal systematics are modeled with $2.0\%$ uncertainty, with the corresponding background uncertainty being $5.0\%$. Similar uncertainties are adopted for the $\overline{\nu}_\mu^{} \rightarrow \overline{\nu}_e^{}$ channel. For the simulation of neutrino events for the $\nu_\mu^{} \rightarrow \nu_\mu^{}$ channel and antineutrino events for the $\overline{\nu}_\mu^{} \rightarrow \overline{\nu}_\mu^{}$ channel, the signal errors are expected to be $5.0\%$ each. The background error for $\nu_\mu^{}$ is also $5\%$. Moreover, there are additional systematic uncertainties expected for $\nu_\tau^{}$ and NC events, which are treated with 20\% and 10\% uncertainties, respectively. The $\nu_\tau^{}$ and NC systematic uncertainties are expected to affect all neutrino oscillation channels.

    In this work, the $\chi^2_{}$ function shown in Eq.~(\ref{eq:chi2}) is computed for each neutrino oscillation channel. The simulated events are summed for signal and background events. The $\chi^2_{}$ function is minimized for the neutrino oscillation parameters $\theta_{13}^{}, \theta_{23}^{}, \delta_{\rm CP}^{}$ and $\Delta m_{31}^2$. The minimization is carried out with the standard minimization algorithm in \texttt{GLoBES}. Parameters $\theta_{12}$ and $\Delta m_{21}^2$ are fixed at their best-fit values, which are expected to be determined at very high precision by solar neutrino experiments and also by the reactor neutrino experiment JUNO~\cite{JUNO:2022mxj}. Unless otherwise stated, it is assumed in our analysis that the octant of $\theta_{23}^{}$ is not known and the true neutrino mass ordering is NO. For the true values for the neutrino oscillation parameters, we adopt the current best-fit results from \texttt{NuFIT 6.0}. For convenience, the best-fit values are summarized in Table~\ref{tab:bestfits}. In order to provide general results, no priors are used for the neutrino oscillation parameters in our numerical analysis. However, to account for the necessary numerical uncertainties that are related to the matter density profile, we incorporate a conservative $2.0\%$ uncertainty ($1\sigma$~CL) for the Shen-Ritzwoller profile, which has been adopted from Ref.~\cite{Roe:2017zdw}. In some instances, one of the NSI parameters is also included in the $\chi_{}^2$ minimization. In those cases, we assume the true value of the NSI parameter to be zero. We also assume no priors for the NSI parameters.

    \begin{table}[!t]
    \begin{center}
        \begin{tabular}{ccc}\hline
        {\bf Parameter} & {\bf Best-fit (NO)} & {\bf Best-fit (IO)} \\ \hline
        \rule{0pt}{3ex}$\sin^2 \theta_{12}^{}$ & 0.308$_{\rm -0.011}^{\rm +0.012}$ & 0.308$_{\rm -0.011}^{\rm +0.012}$ \\ 
        \rule{0pt}{3ex}$\sin^2 \theta_{13}^{}$ & 0.02215$_{\rm -0.00058}^{\rm +0.00056}$ & 0.02236$_{\rm -0.00056}^{\rm +0.00056}$ \\ 
        \rule{0pt}{3ex}$\sin^2 \theta_{23}^{}$ & 0.470$_{\rm -0.013}^{\rm +0.017}$ & 0.550$_{\rm -0.015}^{\rm +0.012}$ \\ 
        \rule{0pt}{3ex}$\delta_{\rm CP}^{}$~[$^\circ$] & 212$_{\rm -41}^{\rm +26}$ & 274$_{\rm -25}^{\rm +22}$ \\ 
        \rule{0pt}{3ex}$\Delta m_{21}^2$~[$10^{-5}~{\rm eV}^2$] & 7.49$_{\rm -0.19}^{\rm +0.19}$  & 7.49$_{\rm -0.19}^{\rm +0.19}$ \\ 
        \rule{0pt}{3ex}$\Delta m_{31}^2$~[$10^{-3}~{\rm eV}^2$] & 2.513$_{\rm -0.019}^{\rm +0.021}$ & $-2.484_{\rm -0.020}^{\rm +0.020}$ \\ \hline
    \end{tabular}
    \end{center}
        \caption{\label{tab:bestfits} The best-fit values of the neutrino oscillation parameters as determined by neutrino oscillation experiments~\cite{Esteban:2024eli}. The uncertainties are presented at $1\sigma$~CL. The values are based on \texttt{NuFIT 6.0}~\cite{NuFIT:6.0}.}
    \end{table}
    
    \section{Numerical Results}

    \label{sec:results}

    In this section, we investigate numerically how one-loop corrections in the matter potential affect the experimental prospects to probe NSIs in neutrino propagation. The investigation is done for the DUNE configuration~\cite{DUNE:2020ypp,DUNE:2021cuw} using the Shen-Ritzwoller profile~\cite{Shen:2016kxw,Roe:2017zdw}.
    
    One-loop effects have two obvious consequences on the search for NSIs in neutrino propagation. First, one-loop corrections in neutrino matter potential could be misidentified as NSIs or any other form of BSM physics that affects neutrino propagation. This is owed to the fact that one-loop effects may lead to a change in neutrino oscillation probability that can be reproduced with non-zero values of one or more NSI parameters. Second, taking one-loop corrections into account in numerical analyses could potentially increase sensitivity to BSM physics as a result of enhanced matter potential. As we will show in the following discussion, both scenarios are viable for NSIs.

    \begin{figure}[!t]
        \centering
        \includegraphics[width=1.0\linewidth]{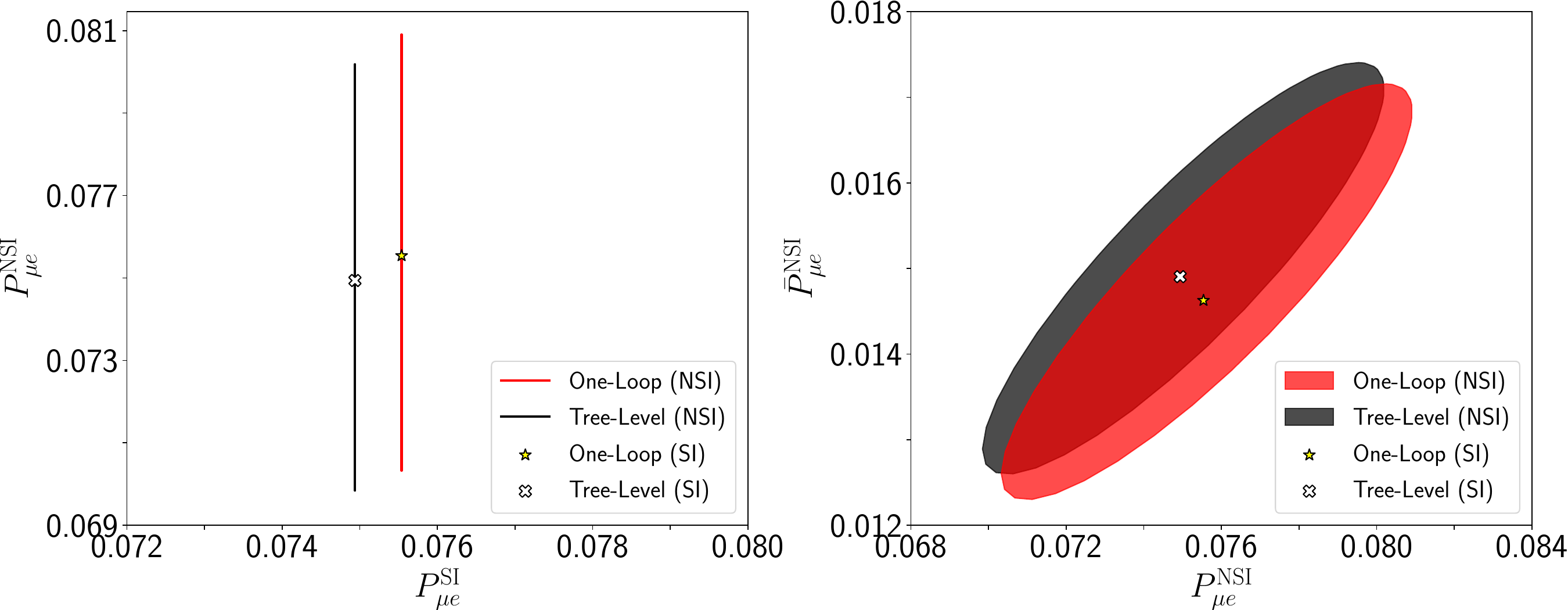}
        \caption{Correlation between the probabilities $P_{\mu e}^{\rm SI}$, $P_{\mu e}^{\rm NSI}$ and $\overline{P}_{\mu e}^{\rm NSI}$. The effect of the NSI parameter $\epsilon_{e\mu}^{}$ is shown by varying $|\epsilon_{e \mu}^{}|$ and $\phi_{e \mu}^{}$ within the ranges $[0, 0.03]$ and $[0, 2\pi]$, respectively. The correlations are shown for $P_{\mu e}^{\rm SI}$ and $P_{\mu e}^{\rm NSI}$ (left panel) and for $P_{\mu e}^{\rm NSI}$ and $\overline P_{\mu e}^{\rm NSI}$ (right panel). The probabilities are computed both at tree-level and including one-loop corrections for the neutrino energy $E_\nu^{} = 2$~GeV. Normal ordering is assumed.}
        \label{fig:biprobability}
    \end{figure}

    In order to find whether one-loop effects can be misidentified as BSM signals, we investigate the correlation between the NSI parameters and one-loop corrections at probability level. In the left panel of Figure~\ref{fig:biprobability}, the correlation between the probabilities $P_{\mu e}^{\rm SI}$ (where the abbreviation SI refers to `standard interaction') and $P_{\mu e}^{\rm NSI}$ is presented at tree-level (black line) and with one-loop corrections (red line). The probabilities are computed for the neutrino energy $E_\nu^{} = 2$~GeV and the DUNE configuration. To account for NSIs, $P_{\mu e}^{\rm NSI}$ is computed by varying the NSI parameter $\epsilon_{e\mu}^{}$ over $|\epsilon_{e \mu}^{}| \in [0, 0.03]$ and $\phi_{e \mu}^{} \in [0, 2\pi]$. Moreover, the cases where $\epsilon_{e\mu}^{} = 0$ for $P_{\mu e}^{\rm NSI}$ are indicated both at tree-level (white cross) and with one-loop corrections (yellow star). For those cases, the probabilities $P_{\mu e}^{\rm NSI}$ and $P_{\mu e}^{\rm SI}$ are equal. In the right panel of Figure~\ref{fig:biprobability}, the correlations are shown for $P_{\mu e}^{\rm NSI}$ and $\overline P_{\mu e}^{\rm NSI}$ at tree-level (black region) and with one-loop corrections (red region). Due to the complexity of $\epsilon_{e\mu}^{}$, both regions appear as ellipses.

    It can be seen in the left panel of Figure~\ref{fig:biprobability} that including one-loop corrections in the neutrino matter potential leads to an increase in $P_{\mu e}^{\rm NSI}$, including the value where $\epsilon_{e\mu}^{} = 0$. By choosing the value of $\epsilon_{e\mu}^{}$ appropriately, the SI probability $P_{\mu e}^{\rm NSI} = P_{\mu e}^{\rm SI}$ could be recovered. Correspondingly, the right panel of the same figure shows that the probabilities computed at tree-level and with one-loop corrections partially overlap when $|\epsilon_{e\mu}^{}|$ and $\phi_{e\mu}^{}$ are allowed to vary. The overlapping area would become larger if $|\epsilon_{e\mu}^{}|$ were allowed to acquire larger values.

\begin{figure}[!t]
        \centering
        \includegraphics[width=1.0\linewidth]{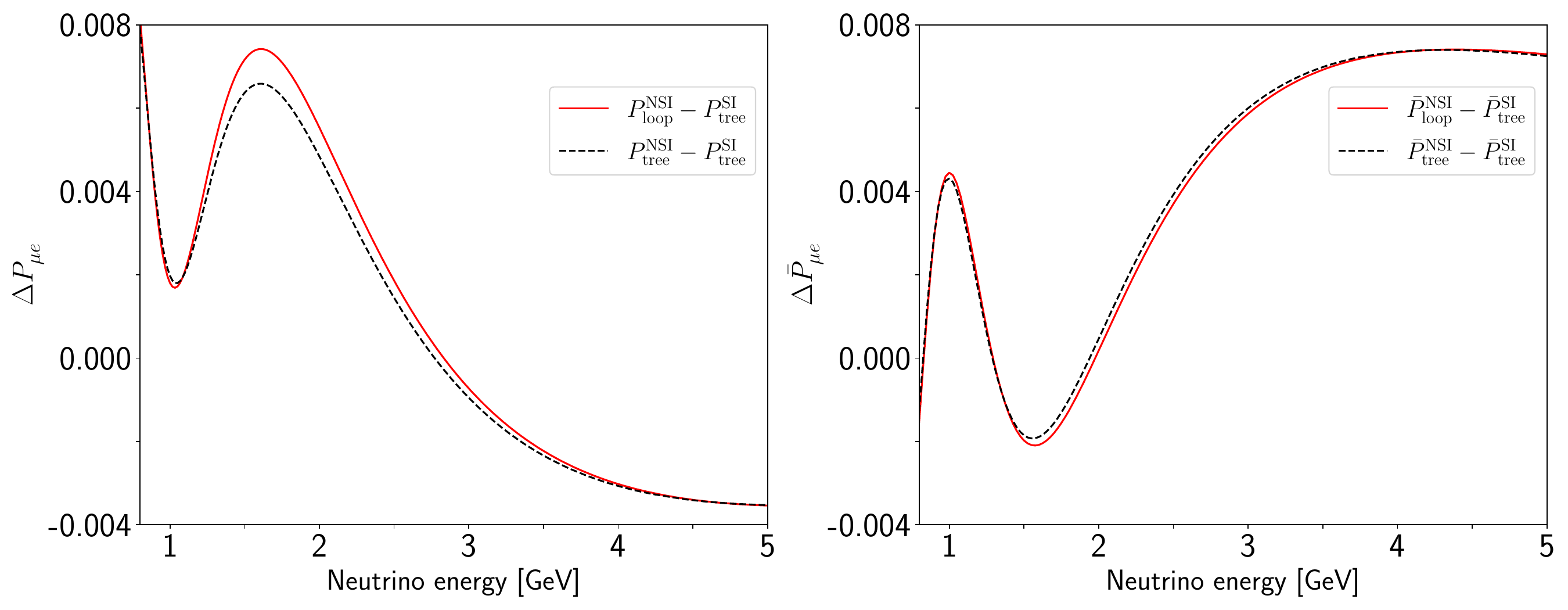}
        \caption{Effect of one-loop corrections on the difference between the neutrino oscillation probability $P_{\mu e}^{}$ computed for the NSI case and the SI case (left panel). The corresponding difference is also shown for the antineutrino probability $\overline{P}_{\mu e}^{}$ (right panel). For the NSI case, we have assumed $\epsilon_{e\mu} = 0.05$. The probability differences are shown both at tree-level (dashed black curve) and with one-loop corrections (solid red curve). Normal ordering is assumed.}
        \label{fig:prob1}
    \end{figure}

    It was shown in our previous work in Ref.~\cite{Huang:2025apv} that one-loop effects lead to an enhancement in the sensitivity to neutrino mass ordering. Similar effects could be expected for the sensitivities to NSIs in neutrino propagation. This effect is shown in Figure~\ref{fig:prob1}, where the differences in the neutrino oscillation probability $P_{\mu e}^{}$ are shown for the case where the NSI effects are present, $P_{}^{\rm NSI} \equiv P_{\mu e}^{} (\epsilon_{e\mu} \neq 0)$, and for the case where only SM interactions are taken into account, $P_{}^{\rm SI} \equiv P_{\mu e}^{}(\epsilon_{e\mu} = 0)$. In Figure~\ref{fig:prob1}, the probability differences are shown for $\nu_\mu^{} \rightarrow \nu_e^{}$ (left panel) and $\overline{\nu}_\mu^{} \rightarrow \overline{\nu}_e^{}$ (right panel) assuming $\epsilon_{e\mu} = 0.05$ in the NSI case. In both panels, the probability differences are presented for neutrino energies $E_\nu^{} \in [1, 5]$~GeV at tree-level (dashed black curves) and with one-loop corrections (solid red curves). Following the methodology presented in Ref.~\cite{Huang:2025apv}, the one-loop corrections to the matter potential is taken to be $\delta a \approx$ 2.0\%. It is observed that the probability differences $\Delta P_{\mu e}^{}$ and $\Delta \overline{P}_{\mu e}^{}$ peak at about $E_\nu^{} =$ 1.8~GeV and 4.0~GeV, respectively. The one-loop effects increase the probability differences $\Delta P_{\mu e}^{}$ for the majority of the displayed neutrino energies. Correspondingly, for antineutrinos the probability differences $\Delta \overline{P}_{\mu e}^{}$ decrease in comparison to the differences that are obtained at tree-level. The figure displays how the one-loop effects would enhance the impact of the NSI parameter $\epsilon_{e\mu}$ on the probabilities.

\begin{figure}[!t]
        \centering
        \includegraphics[width=1.0\linewidth]{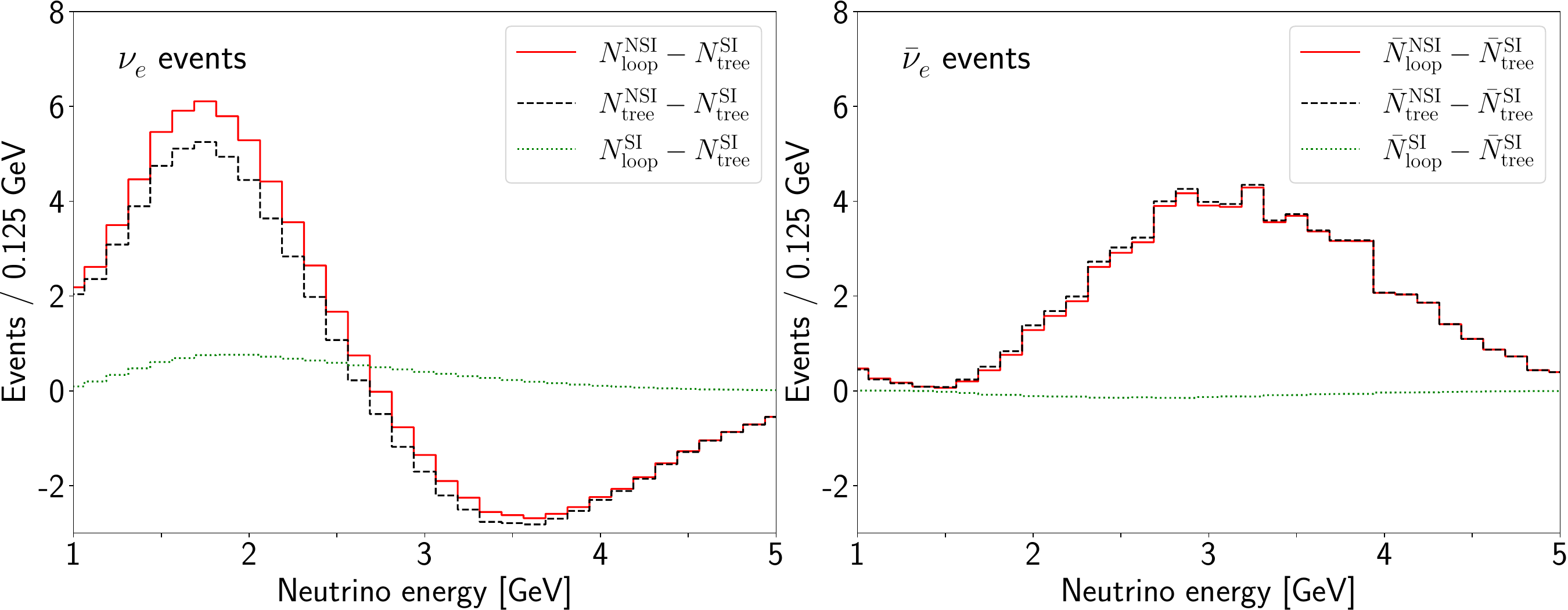}
        \caption{Differences in the expected number of binned $\nu_{e}^{}$ events (left panel) and $\overline{\nu}_{e}^{}$ events (right panel) as a function of neutrino energy. The events are computed in presence of NSIs and one-loop corrections for DUNE. The NSI effects are obtained for $\epsilon_{e\mu} = 0.05$. Normal ordering is assumed.}
        \label{fig:events}
    \end{figure}

    In Figure~\ref{fig:events}, the effect of the NSI parameter $\epsilon_{e\mu}^{}$ and one-loop corrections in the matter potential are illustrated at neutrino event level, assuming NO for the neutrino masses. In the left panel, the difference between the expected $\nu_e^{}$ events obtained with NSIs and without NSIs is shown as a function of neutrino energy for the DUNE configuration and neutrino energies $E_\nu^{} \in [1, 5]$~GeV. In this case, $\nu_e^{}$ and $\overline{\nu}_e^{}$ events $N$ and $\overline{N}$ are obtained at one-loop and tree-level for both NSIs and SIs. To present a concrete example, we have furthermore set $\epsilon_{e\mu}^{} = 0.05$, whereas for the SI case all NSI parameters are set to zero. The dashed green lines depict the difference where one-loop and tree-level events are shown for the SI case. In that case, the change in the expected numbers of $\nu_e^{}$ and $\overline{\nu}_e^{}$ is caused solely by the one-loop corrections, which are mostly positive for $\nu_e^{}$ events and negative for $\overline{\nu}_e^{}$. The effect would be the opposite if the neutrino mass ordering were IO. When the NSI effects are taken into account, the event rates are altered further. This behavior is indicated by the dashed black lines, which illustrate the difference in the expected events between the NSI and SI cases at tree-level. It is evident that the NSI parameter $\epsilon_{e\mu}^{}$ has a stronger effect on the appearance of $\nu_e^{}$ and $\overline{\nu}_e^{}$ for DUNE. Finally, the solid red lines represent the changes in the expected event spectra in the case where the $\nu_e^{}$ and $\overline{\nu}_e^{}$ events are obtained at one-loop for the NSI case and at tree-level for the SI case. The results illustrate how the number of $\nu_e^{}$ and $\overline{\nu}_e^{}$ events, both binned according to the DUNE setup, would change in comparison to the expectations that are computed at tree-level. Therefore, the event differences in Figure~\ref{fig:events} show that the one-loop effects in the matter potential enhances the signal that would be prompted by the NSI parameter $\epsilon_{e\mu}^{}$. The other NSI parameters lead to analogous results.

    The neutrino events that are presented in Figure~\ref{fig:events} are primarily affected by the uncertainty relating to the matter density profile. In the present work, the simulations of DUNE are carried out with the Shen-Ritzwoller profile, which is known to describe the DUNE baseline down to 1\% uncertainty ($1\sigma$~CL)~\cite{Roe:2017zdw}. For such an uncertainty, the effects arising from one-loop corrections stand out from the statistical uncertainties that arise from the matter density profile. Even for a more conservative assumption of 2\%, the effects resulting from the one-loop corrections could produce a noticeable difference, which was also pointed out in our previous study in Ref.~\cite{Huang:2025apv}.

    We next assess the potential effects of including one-loop corrections in neutrino physics analyses. Employing the $\chi_{}^2$ calculation techniques and the DUNE configuration as discussed in Section~\ref{sec:methods}, we calculate the experimental sensitivities that can be expected for DUNE in presence of one-loop corrections. In the following, we provide numerical estimates both on how the one-loop corrections mimic the effect of the NSI parameters and how one-loop corrections can ultimately enhance the sensitivities to the NSI parameters and neutrino mass ordering.
    
\begin{figure}[!t]
        \centering
        \includegraphics[width=1.0\linewidth]{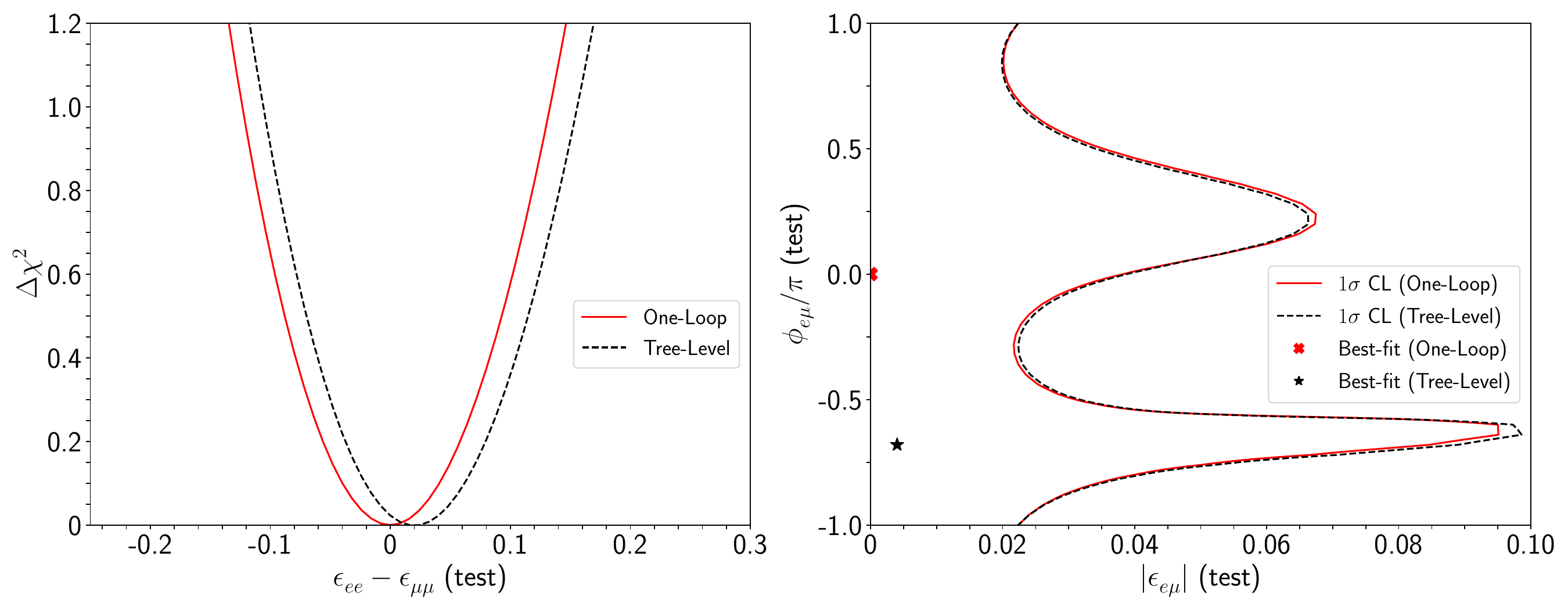}
        \caption{Effect of the one-loop corrections in neutrino matter potential on the fit result for the NSI parameters $\epsilon_{ee}^{} - \epsilon_{\mu\mu}^{}$ (left panel) and $\epsilon_{e\mu}^{}$ (right panel). While the true events are generated with one-loop corrections, the fitted events are computed both at tree-level (dashed black curves) and with one-loop corrections (solid red curves). Normal ordering is assumed.}
        \label{fig:mimick1}
    \end{figure}

    Since one-loop effects are typically not considered in neutrino physics analyses, their presence can be misidentified as a BSM signal. This mimicking effect is illustrated in Figure~\ref{fig:mimick1}. The dashed black lines show the scenario where neutrino events computed for tree-level are fitted as function of test values $\epsilon_{ee}^{} - \epsilon_{\mu\mu}^{}$ (left panel) and $\epsilon_{e\mu}^{}$ (right panel). For concreteness, the neutrino mass ordering is assumed to be NO and the uncertainties related to the matter density profile are omitted. In the left panel, the expected $\Delta \chi_{}^2 \equiv \chi_{\rm NSI}^2 - \chi_{\rm SI}^2$ distribution is shown for tree-level (dashed black curve) and one-loop (solid red curve) as a function of $\epsilon_{ee}^{} - \epsilon_{\mu\mu}^{}$. In the right panel, the expected $1\sigma$~CL contours are shown for both the tree-level case and the one-loop case as functions of the test values of $|\epsilon_{e\mu}^{}|$ and $\phi_{e\mu}^{}$. In both panels, the true value of the respective NSI parameter is zero. In the former case, the best-fit point, which gives the lowest $\chi^2_{}$ value in the fit, is located at the point where the NSI parameters are zero. However, when the test events are obtained at one-loop instead of tree-level, the change in the matter effects indicate a non-zero value of the considered NSI parameter. Therefore, Figure~\ref{fig:mimick1} illustrates how the one-loop effects can mimic the experimental signature that could otherwise indicate existence of BSM physics. In this regard, we find that one-loop corrections could be misinterpreted as a non-zero NSI parameter with the best-fit value $\epsilon_{ee}^{} - \epsilon_{\mu\mu}^{} = 0.02$, or $|\epsilon_{e\mu}^{}| = 0.004$ and $\phi_{e\mu}^{} = -0.68\pi$. In similar manner, we find the best-fit values for the other NSI parameters to be $|\epsilon_{e\tau}^{}| = 0.002$ and $ \phi_{e\tau}^{} = 0.72\pi$, $|\epsilon_{\mu\tau}^{}| = 0.003$ and $\phi_{\mu\tau}^{} = 0.52\pi$, and finally $|\epsilon_{\tau \tau}^{} - \epsilon_{\mu \mu}^{}| = -0.006$.

\begin{figure}[!t]
        \centering
        \includegraphics[width=1.0\linewidth]{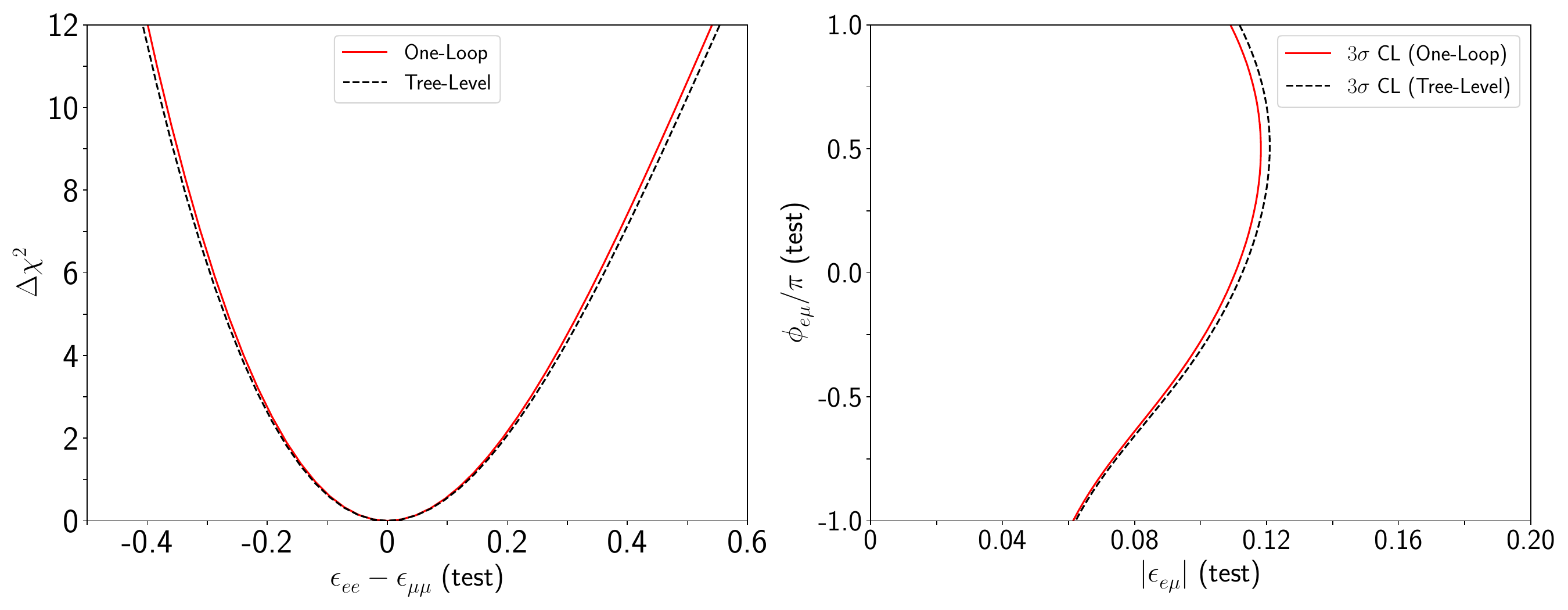}
        \caption{Effect of one-loop corrections on the allowed values for the NSI parameters $\epsilon_{ee}^{} - \epsilon_{\mu\mu}^{}$ and $\epsilon_{e\mu}^{}$. The $\Delta \chi^2 = \chi^2_{\rm NSI} - \chi^2_{\rm SI}$ distributions are shown for both the tree-level and one-loop contributions, respectively. True and test events are both computed with same matter potential, which is obtained at tree-level (dashed black curves) and with one-loop corrections (solid red curves). The true mass ordering is assumed to be normal ordering.}
        \label{fig:nsi_bounds1}
    \end{figure}

    The effect of one-loop corrections on the sensitivities to individual NSI parameters can be computed in a similar manner. In Figure~\ref{fig:nsi_bounds1}, the sensitivities are presented to the NSI parameters $\epsilon_{ee}^{} - \epsilon_{\mu\mu}^{}$ (left panel) and $\epsilon_{e\mu}^{}$ (right panel). In both panels, $\Delta \chi_{}^2 \ \equiv \chi_{\rm NSI}^2 - \chi_{\rm SI}^2$ is shown as a function of the test value of the depicted NSI parameter, while the true value of the same NSI parameter is taken to be zero. The uncertainties related to the matter density profile are taken into account. Meanwhile, in the case of the off-diagonal NSI parameter $\epsilon_{e\mu}^{}$, the complex phase $\phi_{e\mu}^{}$ is allowed vary freely. As before, the sensitivities that are obtained for tree-level are shown by the dashed black curves, while the sensitivities that are obtained for one-loop are indicated by the solid red curves. The neutrino mass ordering is assumed to be NO. The figure displays the allowed values for each NSI parameter at a given CL when the remaining NSI parameters are zero. Analogous results can be found for the remaining NSI parameters.
    
    It is observed in Figure~\ref{fig:nsi_bounds1} that the sensitivities to the NSI parameters $\epsilon_{ee}^{} - \epsilon_{\mu\mu}^{}$ are slightly improved when the one-loop corrections in the matter potential are included. At $3\sigma$~CL, the solid red curves show improved sensitivities in comparison to the sensitivities that are obtained at tree-level. At $90\%$~CL, the improvement due to one-loop effects is less significant, as can be seen in the left panel of Figure~\ref{fig:nsi_bounds1}. One can expect similar enhancements for the IO case.
    
\begin{figure}[!t]
        \centering
        \includegraphics[width=1.0\linewidth]{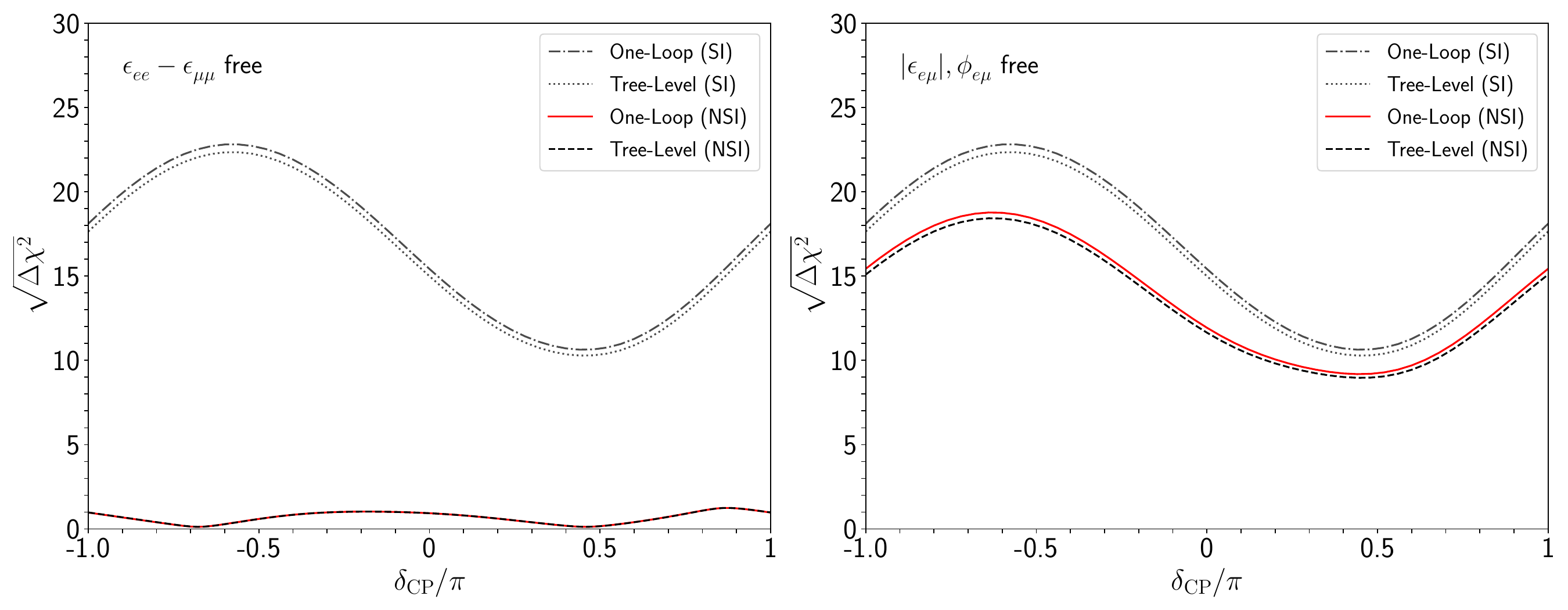}
        \caption{Effect of the NSI parameters $\epsilon_{ee}^{} - \epsilon_{\mu\mu}^{}$ (left panel) and $\epsilon_{e\mu}^{}$ (right panel) on the sensitivity to neutrino mass ordering for DUNE. The sensitivity is shown both at tree-level (dashed black curve) and including one-loop corrections (solid red curve), while assuming normal ordering. In the NSI case, the minimization includes either $\epsilon_{ee}^{} - \epsilon_{\mu\mu}^{}$ or $|\epsilon_{e\mu}^{}|$ and $\phi_{e\mu}^{}$. In the SI case, the NSI parameters are fixed at zero. The sensitivity is also presented for the SI case at tree-level (dotted gray curve) and with one-loop effects (dot-dashed gray curve).}
        \label{fig:mass_ordering}
    \end{figure}

    We finally examine the interplay between the NSI parameters and one-loop corrections in the determination of the neutrino mass ordering. To do this, the sensitivity to neutrino mass ordering is calculated as a function of true value of $\delta_{\rm CP}^{}$ for the NSI parameters $\epsilon_{e\mu}^{}, \epsilon_{e\tau}^{}, \epsilon_{\mu\tau}^{}, \epsilon_{ee}^{} - \epsilon_{\mu\mu}^{}$ and $\epsilon_{\tau\tau}^{} - \epsilon_{\mu\mu}^{}$. The sensitivity to rule out the wrong mass ordering is obtained as $\sqrt{\Delta \chi_{}^2} \equiv \sqrt{\chi_{\rm IO}^2 - \chi_{\rm NO}^2}$, where the minimization of $\chi_{}^2$ is done for the indicated NSI parameter in addition to the standard neutrino oscillation parameters. Moreover, $\chi_{\rm IO}^2$ is minimized over the test values of $\Delta m_{31}^2$ that correspond to IO, whilst $\chi_{\rm NO}^2$ is minimized for the test values that correspond to NO. The minimization is done separately for $\chi_{\rm NO}^2$ and $\chi_{\rm IO}^2$. The results are exemplified in Figure~\ref{fig:mass_ordering}. The uncertainties related to the matter density profile are taken into account.
    
    In Figure~\ref{fig:mass_ordering}, the sensitivities to neutrino mass ordering are shown for tree-level (dashed black curves) and one-loop level (solid red curves) when the depicted NSI parameter is allowed to vary freely. As an example, we present the sensitivity to neutrino mass ordering for $\epsilon_{ee}^{} - \epsilon_{\mu\mu}^{}$ (left panel) and $\epsilon_{e\mu}^{}$ (right panel). Additionally, the sensitivities are shown for the cases where all of the NSI parameters are assumed to be zero. Those sensitivities are shown for tree-level (dot-dashed gray curves) and one-loop (dotted gray curves). It is evident from Figure~\ref{fig:mass_ordering} that letting any of the NSI parameter vary freely leads to a notable drop in the sensitivity. For example, letting $\epsilon_{e\mu}$ to vary freely in the minimization reduces the sensitivity at one-loop level from $22.2\sigma$~CL to $18.0\sigma$~CL at tree-level for $\delta_{\rm CP} = -\pi/2$, indicating a reduction of $4.2\sigma$~CL. On the other hand, the sensitivities are slightly recovered when the one-loop effects are taken into account. For the same benchmark value, $\delta_{\rm CP} = -\pi/2$, the improvement is about $0.3\sigma$~CL. One can therefore infer that the effect of the NSI parameters is more significant than the enhancement that can be expected from one-loop corrections. However, it is revealed in Figure~\ref{fig:mass_ordering} that DUNE can reach the $5\sigma$~CL limit despite the effects of $\epsilon_{e\mu}$. We compute the sensitivities to neutrino mass ordering similarly for other NSI parameters. We find that the sensitivity to neutrino mass ordering stays above the critical $5\sigma$ CL benchmark when either $\epsilon_{\mu\tau}^{}$ or $\epsilon_{\tau\tau}^{} - \epsilon_{\mu\mu}^{}$ are allowed to run free in the $\chi^2_{}$ fits. In contrast, letting $\epsilon_{e\tau}^{}$ or $\epsilon_{ee}^{} - \epsilon_{\mu\mu}^{}$ vary freely would reduce the sensitivity to neutrino mass ordering to about $2\sigma$ and $1\sigma$~CL, respectively. For those two cases, including one-loop corrections in the matter potential is not adequate to restore the sensitivity to $5\sigma$ or even $3\sigma$~CL.

    We finally note that the data used to create Figures~\ref{fig:prob2}--\ref{fig:mass_ordering} as well as other numerical results in this article is publicly available in Ref.~\cite{OurData}.
    
	\section{Summary}
	
	\label{sec:sum}
	
	In this work, we investigate the effects of one-loop corrections to the neutrino matter potential when NSIs are introduced for neutrino oscillations in DUNE. With a $2.0\%$ correction to the CC matter potential, we examine for the first time how it affects the experimental constraints on the NSI parameters and the physics goals of DUNE. In the assumption of the NSI parameter $\epsilon_{e \mu}^{} > 0$, numerical simulations in the appearance channel $\nu_{\mu}^{} \to \nu_{e}^{}$ reveal that the one-loop effect increases the difference in the numbers of $\nu_e^{}$ events between NSIs and SIs in the NO case, while such a difference decreases for antineutrino events.
	
	Omitting the one-loop effects could lead to wrong BSM signals. We have found that incorrect best-fit values of the NSI parameters will be obtained when fitting the experimental data with the tree-level matter potential. This indicates that the mimicking effects between NSIs and the one-loop matter potential will influence our analyses of BSM physics. 
	
	Finally, the NSI effects would generally reduce the sensitivities to neutrino mass ordering in DUNE. With the benchmark value $\delta_{\rm CP}^{} = -\pi/2$, there is a reduction of $4.2\sigma$ CL for freely-varying $\epsilon_{e \mu}^{}$ in the NO case, while the one-loop effects slightly increase the sensitivities by about $0.3\sigma$ CL. However, the situation will become worse when considering the variation of $\epsilon_{ee}^{}-\epsilon_{\mu\mu}^{}$, and the total sensitivities will drop to less than $3\sigma$ CL. In conclusion, one-loop effects from the standard electroweak interactions should be incorporated consistently in future long-baseline neutrino experiments for both precise measurements of oscillation parameters and studies on new physics beyond the SM.

	\section*{Acknowledgements}
	
	This work was supported in part by the National Natural Science Foundation of China under grant No.~12475113, by the CAS Project for Young Scientists in Basic Research (YSBR-099), and by the Scientific and Technological Innovation Program of IHEP under grant No.~E55457U2.

	\bibliographystyle{elsarticle-num}
	\bibliography{References}

\end{document}